 \documentclass[12pt]{iopart}
 \input{epsf.sty}
 \input{epsf.tex}
 \newcommand{\be}{\begin{equation}}
 \newcommand{\ee}{\end{equation}}
 \newcommand{\ba}{\begin{eqnarray}}
 \newcommand{\ea}{\end{eqnarray}}
 \makeatletter
 \@addtoreset{equation}{section}
 \makeatother
 \renewcommand{\theequation}{\thesection.\arabic{equation}}
 \def\appendix #1#2
 {
  \section*{Appendix #1: #2}
  \renewcommand{\theequation}{#1.\arabic{equation}}
 }
 \def\Haj{H\'aj\'\i\v cek~}
 
 \def\Schr{Sch\-r\"o\-din\-ger}
 \def\sinh{\hbox{\rm sinh}}
 \def\cosh{\hbox{\rm cosh}}
 \def\tanh{\hbox{\rm tanh}}
 \def\coth{\hbox{\rm coth}}

 \def\arcoth{\hbox{\rm arcoth}}
 \def\d{\partial}
 \def\h{\hat}
 \def\ra{\rightarrow}
 
 \def\pre{PRBB}
 \def\post{POBB}
 \def\real{{\vrule height 1.6ex
            width 0.05em depth 0ex \kern -0.06em {\rm R}}}
 \def\entry#1#2{\vbox{\hbox to 112truept{\hrulefill}\break%
                \hbox{\vrule\vbox to 28truept{%
                \vfill%
                \hbox to 112truept{\hfill\quad\small #1\quad\hfill}\break%
                \vfill%
                \hbox to 112truept{\hfill\quad\small #2\quad\hfill}%
                \break\vfill%
                \hbox to 112truept{\hrulefill}}\vrule}}}%
 \def\arrwv#1#2{\vbox to 45truept{\vfill%
               \hbox to 112truept{\hfill
               \put(42,20){\small#1}
               \put(56,40){\vector(0,-1){40}}
               \put(62,20){\small#2}\hfill}
               \vfill}}
 \def\arrwhdd#1#2{\vbox to 28truept{\vfill
                \hbox to 92truept{\hfill
                \put(44,2){\small#1}
                \put(4,14){\vector(1,0){86}}
                \put(44,18){\small#2}\hfill}
                \vfill}}
 \def\cross#1#2{\vbox to 45truept{\vfill%
                \hbox to 112truept{\hfill
                \put(0,32){\small#1}
                \put(14,42){\vector(2,-1){86}}
                \put(102,32){\small#2}
                \put(14,0){\vector(2,1){86}}\hfill}
                \vfill}}
 \def\sc{\scriptstyle}
 
 \def\ASP{{\it Astropart.\ Physics~}}

 \def\IJMP{{\it Int.\ J.\ Mod.\ Phys.\ ~}}
 \def\MPL{{\it Mod.\ Phys.\ Lett.\ ~}}
 \def\GRG{{\it Gen.\ Rel. Grav.~}}
 
\begin{document}

\title{Canonical and path integral quantisation of string cosmology models}

\author{Marco Cavagli\`a\footnote[1]{E-mail: cavaglia@aei-potsdam.mpg.de.} 
and Carlo Ungarelli\footnote[2]{E-mail: ungarell@aei-potsdam.mpg.de.}}

\address{Max-Planck-Institut f\"ur Gravitationsphysik,
Albert-Einstein-Institut, Schlaatzweg 1, D-14473 Potsdam, Germany}

\begin{abstract}
We discuss the quantisation of a class of string cosmology models that are 
characterized by scale factor duality invariance. We compute the amplitudes for 
the full set of classically allowed and forbidden transitions by applying the 
reduce phase space and the path integral methods. We show that these approaches 
are consistent. The path integral calculation clarifies the meaning of the 
instanton-like behaviour of the transition amplitudes that has been first 
pointed out in previous investigations. 
\end{abstract}

\pacs{04.60.Kz,98.80.Cq,98.80.Hw}



\section{Introduction}
String theory, thanks to its duality symmetries, provides a cosmological 
scenario \cite{GV0,GV1,GV2} in which the Universe starts from the perturbative 
vacuum of (super)string theory and evolves in a `pre-big bang' (\pre) phase 
\cite{GV0,GV1} characterized by an accelerated growth of the curvature and of 
the string coupling.

One of the main problems of string cosmology is the understanding of the 
mechanism responsible for the transition (`graceful exit') from the 
inflationary \pre\ phase to the deflationary `post-big bang' phase (\post) with 
decreasing curvature that is typical of the standard cosmological scenario. 
Necessarily, the graceful exit involves a high-curvature, strong coupling, 
regime where higher derivatives \cite{GMaV} and string loops terms must be 
taken into account. In \cite{Grex} it has been shown that for any choice of the 
(local) dilaton potential no cosmological solutions that connect smoothly the 
\pre\ and \post\ phases do exist. As a consequence, at the classical level 
higher order corrections cannot be `simulated' by any realistic dilaton 
potential. 

At the quantum level the dilaton potential may induce the transition from the 
\pre\ phase to the \post\ phase. In this context, using the standard Dirac 
method of quantisation based on the Wheeler-De Witt equation \cite{WDW} a 
number of minisuperspace models have been investigated in the literature 
\cite{GV,GVM,Mah}. The result of these investigations is a finite, non-zero, 
transition probability \pre\ $\rightarrow$ \post\ with a typical 
`instanton-like' dependence ($\sim\,\exp\{-1/g^2\}$) on the string coupling 
constant \cite{GV,GVM}.

The aim of this paper is to present a refined analysis of the quantisation of 
string cosmological models. To this purpose we reconsider the minisuperspace 
models that have been previously investigated in \cite{GV,GVM,Mah}. We have 
several motivations for doing this.  

First of all, these systems are invariant under reparametrisation of time.  So
their quantisation requires a careful discussion of the subtleties that are
typical of the quantisation of gauge invariant systems (e.g.\ gauge fixing)
\cite{CD,time}. Furthermore, we want to investigate the graceful exit in string
cosmology using different techniques of quantisation and illustrate a consistent
approach to the problem that can be successfully applied to a large class of
models.

We deal with a class of string inspired models -- see (\ref{action-can}) and
(\ref{H}) below -- that are exactly integrable and we apply the standard
techniques for the canonical quantisation of constrained systems
\cite{dirac64,HRT,HT}. Using the reduced-phase space formalism we determine the
positive norm Hilbert space of states. We construct the \pre\ and \post\ wave
functions that are normalized with respect to the inner product of the Hilbert
space. These wave functions are then used to compute transition amplitudes.
Further, we compute the (semiclassical) transition amplitude \pre\ $\rightarrow$
\post\ by the path integral approach. The result agrees with the semiclassical
limit of the transition amplitude that has been obtained in the reduced-phase
space approach and makes clear the instanton-like structure pointed out in
\cite{GV,GVM}. Let us stress that our investigation is important at least for
two reasons: First, the model that we are discussing is (to our knowledge) the
only known example of a minisuperspace model where {\it exact} transition
probabilities between two classically disconnected backgrounds have been
calculated. Second, our analysis completes the previous investigations of
\cite{GV,GVM,Mah} and allows for a systematic discussion of both classically
allowed and classically forbidden transitions.

The outline of the paper is as follows. Sect.\ 2 is devoted to the 
classical theory. We derive the solutions of the equations of motion and 
discuss the classical behaviour of the \pre\ and \post\ branches. In Sect.\ 3 
we quantise the  model. This task is completed using first the canonical 
approach and then the path integral formalism. Eventually, we state our 
conclusions in Sect.\ 4.
\section{Classical theory}
We consider the string inspired model in d+1 dimensions described by the action
(we assume that only the metric and the dilaton contribute non-trivially to the 
background) 
\be
S = \frac{1}{2\,\lambda_s^{{\rm d}-1}}\,\int\,d^{{\rm 
d}+1}x\,\sqrt{|g|}\,e^{-\phi}\,\left(R+\partial_{\mu}\phi\partial^{\mu}\phi 
-V(g_{\mu\nu},\phi)\right )\,,
\label{Act1}
\ee
where $\phi$ is the dilaton field, $\lambda_s=(\alpha^{\prime})^{1/2}$ is the
fundamental string length parameter, and $V(g_{\mu\nu},\phi)$ is a potential 
term. When the latter is absent, (\ref{Act1}) coincides with the tree-level, 
lowest order in $\alpha^{\prime}$, string effective action \cite{MT}
defined in the `String Frame', where the metric $g_{\mu\nu}$ coincides with the
$\sigma$-model background metric that couples directly to the strings.

We deal with isotropic, spatially flat, cosmological backgrounds parametrized by 
\be
g_{\mu\nu} ={\rm diag} \left(-N^2(t), a^2(t) \delta_{ij}\right)\,,\quad
a= \exp\left[\Theta (t)/\sqrt{{\rm d}}\right]\,,\quad \phi=\phi(t)\,,
\label{par}
\ee
where $i,j=1,..$d. We also assume that the spatial sections have finite volume. 
For this class of backgrounds the action (\ref{Act1}) reads 
\be
S=\int\,dt\,{\cal L}\,,~~~~~
{\cal L}={\lambda_s\over 2}\, \left({1\over\mu}(\dot{\Theta}^2-\dot{\Phi}^2)-
\mu\,\e^{-2\Phi}V(\Theta,\Phi)\right)\,.\label{new-act}
\ee
where dots represent differentiation with respect to cosmic time
$t$, $\mu(t)=N(t)\,e^{\Phi}\,$, and $\Phi$ is the `shifted' dilaton field
\be
\Phi=\phi-\log\,\int\,d^{\rm d}x/\lambda_s^{\rm d}-\sqrt{{\rm d}}\,\Theta \,.
\label{sdil}
\ee
(In (\ref{new-act}) we have neglected surface terms that are inessential for 
our purposes.) In this paper we restrict attention to models with potential term 
depending on the shifted dilaton $\Phi$ only. In this case (\ref{new-act}) is 
invariant under scale factor duality transformations \cite{GV0,sfd} 
\ba
&& \Phi\ra\Phi~~~{\rm or}~~~\phi\ra\phi-2{\rm d}\,\log a\,,\\
&&\Theta\ra-\Theta~~~{\rm or}~~~a\ra\frac{1}{a}\,.
\ea
Let us introduce the conjugate momenta to $\Theta$ and  $\Phi$
by the Legendre transformation
\be
\Pi_{\Theta}=\frac{\lambda_s}{\mu}\,\dot{\Theta}\,,\qquad
\Pi_{\Phi}=-\frac{\lambda_s}{\mu}\,\dot{\Phi}\,.
\ee
Equation (\ref{new-act}) can be cast in the canonical form
\be
S=\int dt\,\left[\dot\Theta \Pi_\Theta+\dot\Phi \Pi_\Phi-
{\cal H}\right]\,,\label{action-can}
\ee
where 
\be
{\cal H}=\mu(t) H\,,~~~~
H={1\over 2\lambda_s}
\left(\Pi_\Theta^2-\Pi_\Phi^2+\lambda_s^2 V(\Phi)\,e^{-2\Phi}\right)\,.
\label{H}
\ee
In the canonical formalism $\mu$ plays the role of a non-dynamical variable 
that enforces the constraint $H=0$. As we do expect for a 
time-reparametrisation invariant system the total Hamiltonian ${\cal H}$ is 
proportional to the constraint \cite{HRT,HT}. The equations of motion are
\ba
&&{d\Theta\over d\tau}=\frac{\Pi_{\Theta}}{\lambda_s}\,,~~~~~~
{d\Phi\over d\tau}=-\frac{\Pi_{\Phi}}{\lambda_s}\,,\label{eqm1}\\
&& {d\Pi_{\Theta}\over d\tau}=0\,,~~~~~~
{d\Pi_{\Phi}\over d\tau}=
{\lambda_s\over 2}\,\left(2V-\frac{dV}{d\Phi}\right)\,
e^{-2\Phi}\,, 
\label{eqm2}
\ea
where $\tau(t)$ is 
\be
\tau(t)=\int^t ds \, \mu(s)\,.
\ee
The gauge parameter $\tau(t)$ is related to the synchronous-gauge time 
$t^{(sg)}$ ($N=1$) by the relation
\be
t^{(sg)}(\tau)=\int^{\tau} ds\,e^{-\Phi}\,.
\label{tsg}
\ee
We consider potentials of the form
\be
V(\Phi)=\lambda\,e^{-2\Phi(q-1)}\,,
\label{Potq}
\ee
where $\lambda>0$ is a dimension-two quantity (in natural units) and $q$ is a 
dimensionless parameter. (This class of potentials has been first discussed 
in~\cite{Mah}.) For $q\neq 0$ the explicit solution of the equations 
of motion (\ref{eqm1}), (\ref{eqm2}) is
\be
\begin{array}{ll}
\Theta=\displaystyle\Theta_0+\frac{k}{\lambda_s}\,(\tau-\tau_0)\,,\quad&
e^{\Phi}=\displaystyle\left[{\sqrt{\lambda}\lambda_s\over|k|}\,
\sinh\left({|kq|\over\lambda_s}|\tau-\tau_0|\right)\right]^{1/q}\,,\\\\
\Pi_{\Theta}=k\,,&\displaystyle\Pi_{\Phi}=-k\,\coth
\left[{kq\over\lambda_s}(\tau-\tau_0)\right]\,.
\end{array}
\label{sol12}
\ee
(The case $q=0$ corresponds -- modulo a redefinition of $k$ -- to the 
`vacuum' solutions discussed in \cite{GV0,GV1,GV2}.) Let us determine which 
values of $q$ do allow for the existence of an inflationary expanding \pre\ 
branch and a decelerating \post\ branch. According to the general analysis of 
\cite{GV0,GV1}, the expanding \pre\ and \post\ branches are defined by
\ba
&{\rm \pre}:&~~~~{\bf H}>0\,,~~~~\dot{{\bf H}}>0\,,~~~~\dot{\Phi}>0\,,\\
&{\rm \post}:&~~~~{\bf H}>0\,,~~~~\dot{{\bf H}}<0\,,~~~~\dot{\Phi}<0\,,
\ea
where 
\ba
&&{\bf H}=\frac{d~~}{d t^{(sg)}}\,(\log a)=\frac{1}{\sqrt{{\rm d}}}\,
\frac{k}{\lambda_s}\,e^{\Phi}\,,
\label{Hub}\\
&&\dot{{\bf H}}=\frac{d~~}{d t^{(sg)}}\,{\bf 
H}=\frac{1}{\sqrt{{\rm d}}}\,\frac{k}{2\lambda_s}\,
\frac{d~}{d\tau}\,(e^{2\Phi})\,,
\label{dHub}\\
&&\dot\Phi=\frac{d~~}{d t^{(sg)}}\,\Phi=\frac{d}{d\tau}\,(e^{\Phi})\,.
\label{dphi}
\ea
From (\ref{Hub}) it is straightforward to see that expanding and 
contracting backgrounds are identified by $k>0$ and $k<0$ respectively. $k=0$ 
corresponds to the flat (d+1)-dimensional Minkowski space. Since we are 
interested in expanding backgrounds here and throughout the paper we shall 
consider only positive values of $k$, i.e.\ solutions with $\Pi_\Theta>0$. For 
$q\leq 1$ we have two distinct branches corresponding to \pre\ and \post\ 
states. (The limiting case $q=1$ corresponds to a positive constant potential 
in (\ref{Act1}). The relative classical solutions have been discussed in 
\cite{Cosmol}.) The \pre\ $(+)$ and \post\ $(-)$ branches are identified by 
negative and positive values of $\Pi_{\Phi}$ respectively. Asymptotically, for 
$0<q\le 1$ we have
\be
\lim_{\Phi\rightarrow +\infty}\Pi^{(\pm)}_{\Phi}=\mp 
k\,,~~~~~\lim_{\Phi\rightarrow -\infty}\Pi^{(\pm)}_{\Phi}
\sim \mp\lambda_s\sqrt{\lambda}e^{-q\Phi}\,,
\ee
in the strong and weak coupling regime respectively. Conversely, for $q<0$ 
we have
\be
\lim_{\Phi\rightarrow+\infty}\Pi^{(\pm)}_{\Phi}
\sim \mp\lambda_s\sqrt{\lambda}e^{-q\Phi}\,,~~~~~
\lim_{\Phi\rightarrow-\infty}\Pi^{(\pm)}_{\Phi}=\mp k\,.
\label{p2qmi}
\ee
Substituting (\ref{sol12}) in (\ref{tsg}) the synchronous-gauge time $t^{(sg)}$ 
can be written explicitly in terms of $\tau$. We distinguish two different 
cases:

{\it i)} $q\neq\frac{1}{2m+1}$, $m=0,1,2,...$
\ba
t^{(sg)}-t^{(sg)}_0=&&-{\sigma(q)\sigma(\tau-\tau_0)\over
\sqrt{\lambda}|q-1|}
\left(\frac{\sqrt{\lambda}\lambda_s}{k}\right)^{1-1/q}\cdot\nonumber\\
&&\cdot [\,\sinh(x)]^{1-1/q}
F\left({\sc\frac{\sc 1}{\sc 2},\frac{\sc q-1}{\sc 2q},\frac{\sc 3q-1}{\sc 
2q}},-\sinh^2x\right)\,,\nonumber
\ea
where $\sigma$ is the sign function, $F$ is the hypergeometric function 
\cite{AS}, and
$$
x=\frac{k}{\lambda_s}|q(\tau-\tau_0)|\,;
$$

{\it ii)} $q=\frac{1}{2m+1}$, $m=0,1,2,...$ 
$$
t^{(sg)}-t^{(sg)}_0=-{\sigma(\tau-\tau_0)\over 2\sqrt{\lambda}|q|}
\left(\frac{\sqrt{\lambda}\lambda_s}{k}\right)^{1-1/q}
\frac{\Gamma(m+1/2)}{\sqrt{\pi}\Gamma(m+1)}f(m,x)\,,
$$
where
$$
f(0,x)=4\,\arcoth(e^x)\,,
$$
and 
$$
f(m,x)=(-1)^m\,f(0,x)+\cosh(x)\sum_{k=0}^{m-1}
\frac{(-1)^k\Gamma(m-k)\sqrt{\pi}}{\Gamma(m+1/2-k)[\sinh(x)]^{2(m-k)}} 
$$
for $m>0$.
The above relations determine the \pre\ and \post\ branches in terms of the 
synchronous-gauge time  $t^{(sg)}$ for different values of the parameter $q$. In 
particular, we have 

{\it a)} $q<0$.
In this case the \pre\ and \post\ branches are defined for
\be
\begin{array}{rllrll}
-\infty &<\tau-\tau_0,<0&\ra&-\infty &<t^{(sg)}-t^{(sg)}_0&<0\,,\\
0&<\tau-\tau_0<\infty&\ra&0&<t^{(sg)}-t^{(sg)}_0&<\infty\,,
\end{array}
\ee
respectively.

{\it b)} $q=\frac{1}{2m+1}$. 
The \pre\ and the \post\ branches are defined for
\be
\begin{array}{rllrll}
0&< \tau-\tau_0<\infty&\ra&-\infty&<t^{(sg)}-t^{(sg)}_0&<0\,,\\
-\infty&<\tau-\tau_0<0&\ra&0&<t^{(sg)}-t^{(sg)}_0&<\infty\,,
\end{array}
\ee
respectively. This can be checked using the asymptotic expansions of $f(m,x)$ 
for $x\to 0$ and $x\to +\infty$.

{\it c)} $0<q<1$, $q\neq\frac{1}{2m+1}$ .
The \pre\ branch is defined for
\be
0<\tau-\tau_0<\infty~\ra~\infty<t^{(sg)}-t^{(sg)}_0<-T\,,\nonumber
\ee
and the \post\ branch for
\be
-\infty<\tau-\tau_0<0~\ra~T<t^{(sg)}-t^{(sg)}_0<\infty\,,\nonumber
\ee
where
\be
T=\frac{1}{2\sqrt{\pi}|q|}\,
\left(\frac{\sqrt{\lambda}\lambda_s}{k}\right)^{1-1/q}\,
\Gamma\left(\frac{q-1}{2q}\right)\,\Gamma\left(\frac{1}{2q}\right)\,.\nonumber
\ee
As it has been pointed out in \cite{GVM,Mah}, in terms of the synchronous-gauge 
time $t^{(sg)}$ the \pre\ and \post\ branches are separated by a finite interval 
$\Delta t^{(sg)}=2T$. However, the separation between the two branches has not 
physical meaning. Indeed, due to the presence of a singularity in the curvature 
and in the string coupling the \pre\ and \post\ solutions are disjoint. 
Therefore, it is possible to define the initial value 
of 
$t$ such that the singularity occurs at $t^{sg}=0$ in both branches.
\section{Quantum theory}
The string cosmological model of Sect.\ 2 is described by a
time-reparametrisation invariant Hamiltonian system with two degrees of freedom.
Though its quantisation involves subtleties typical of gauge invariant systems
\cite{CD,time,CDF} the standard techniques of quantisation of constrained
systems can be applied straightforwardly, thanks to the integrability properties
of the model \cite{dirac64,HRT,HT}.

The starting point is the canonical action (\ref{action-can}). Since the
constraint $H$ is of the form $H=H_\Theta(\Theta)+H_\Phi(\Phi)$ the time
parameter can be defined by a single degree of freedom. In the previous
section we have seen that the sign of $\Pi_\Theta$ determines the contracting
vs.\ expanding behaviour of the solutions and the sign of $\Pi_\Phi$
identifies the \pre\ vs.\ \post\ phases. Since we are interested in the
calculation of the quantum transition probability from a (expanding) \pre\
phase to a (expanding) \post\ phase, it is natural to use the
$(\Theta,\Pi_\Theta)$ degree of freedom to define the time of the system and
fix the gauge. In this case the eigenstates of the effective Hamiltonian are
identified by a continuous quantum number $k$ corresponding to the classical
value of $\Pi_\Theta$. Wave functions that describe expanding (contracting) 
solutions are eigenstates of the effective Hamiltonian with $k>0$ ($k<0$).

Let us consider the canonical transformation \cite{CD} 
$(\Theta,\Pi_\Theta,\Phi,\Pi_\Phi)\to (\Sigma,\Pi_\Sigma,\Phi,\Pi_\Phi)$ where
\be
\Sigma=\lambda_s{\Theta\over\Pi_\Theta}\,,~~~~~~\Pi_\Sigma={1\over 
2\lambda_s}\Pi_\Theta^2\,.\label{can-tr-qsc}
\ee
In terms of the new canonical variables the constraint (\ref{H}) reads
\be
H(\Sigma,\Pi_\Sigma,\Phi,\Pi_\Phi)=\Pi_\Sigma-{1\over 
2\lambda_s}\left[\Pi_\Phi^2-\lambda\lambda_s^2 
e^{-2q\Phi}\right]=0\,.\label{H-qsc}
\ee
From (\ref{H-qsc}) it is straightforward to see that $\Sigma$ is canonically 
conjugate to $H$. Thus it defines a global time parameter \cite{time,haj}.
In particular, the gauge fixing identity can be chosen as
\be
F(\Sigma;t)\equiv \Sigma+t=0\,.\label{gf-qsc}
\ee
Equation (\ref{gf-qsc}) fixes the Lagrange multiplier as $\mu=-1$. The 
gauge-fixed action reads
\be
S_{\rm eff}=\int_{t_1}^{t_2} dt\,\left[\dot\Phi\,\Pi_\Phi\,-\,H_{\rm 
eff}(\Phi,\Pi_\Phi)\right]\,,
\label{action-eff-qsc}
\ee
where the effective Hamiltonian is
\be 
H_{\rm eff}(\Phi,\Pi_\Phi)={1\over 
2\lambda_s}\left(\Pi_\Phi^2-\lambda\lambda_s^2 
e^{-2q\Phi}\right)\,.\label{H-eff-qsc}
\ee
The system described by the effective Hamiltonian (\ref{H-eff-qsc}) is
free of gauge degrees of freedom and its quantisation can be performed using 
the standard techniques. In the next two subsections we shall discuss the 
reduced phase space and path integral quantisation procedures. 

\subsection{Reduced phase space quantisation}

The reduced phase space is described by a single degree of freedom with 
canonical coordinates $(\Phi\in\real,\Pi_\Phi\in\real)$. Thus there are no 
ambiguities in the choice of the measure in the Hilbert space: $d[\mu]=d\Phi$. 
In the standard operator approach the quantisation of the model is obtained by 
identifying the canonical coordinates with operators. In the \Schr\ 
representation the self-adjoint operators with respect to the measure $d[\mu]$ 
are 
\be
\Phi\to\h\Phi=\Phi\,,~~~~~\Pi_\Phi\to\h\Pi_\Phi=-i{\d~\over\d\Phi}\,.
\label{op-repr}
\ee
Since the effective Hamiltonian is quadratic in the momenta there are 
no factor ordering ambiguities. The \Schr\ equation reads
\be
-i{\d~\over\d t}\,\Psi(\Phi;t)\,=\, {1\over 2\lambda_s}\left[{\d^2~\over 
\d\Phi^2}+\lambda\lambda_s^2 e^{-2q\Phi}\right]\,\Psi(\Phi;t)\,,
\label{schroedinger-qsc}
\ee
where $t$ is defined by (\ref{gf-qsc}). Finally, the inner product in the 
Hilbert space is
\be
\left(\Psi_2,\Psi_1\right)\,=\,\int_{-\infty}^{+\infty}d\Phi\,
\Psi_2^*(\Phi;t)\,\Psi_1(\Phi;t)\,.\label{prod-int}
\ee
The general solution of the \Schr\ equation (\ref{schroedinger-qsc}) can be 
written as
\be
\Psi_{q}(\Phi;t)=\int dk\, 
A(k)\,\psi_{k,q}(\Phi)\,e^{-ik^2t/2{\lambda_s}}\,,
\ee
where $\psi_{k,q}(\Phi)$ is the solution of the stationary \Schr\ equation $\h 
H_{\rm eff}\psi=E\psi$ with energy $E=k^2/2\lambda_s$
\be
\left[{d^2~\over d\Phi^2}+\lambda\lambda_s^2 
e^{-2q\Phi}\right]\,\psi_{k,q}(\Phi)=-k^2\psi_{k,q}(\Phi)\,.
\label{schroedinger-stat-qsc}
\ee
For $q\not=0$ we have 
\be
\psi_{k,q}(z)=A_1(k,q)J_{i\nu}(z)+A_2(k,q)Y_{i\nu}(z)\,,
\label{sol-gen}
\ee
where $A_1(k,q)$ and $A_2(k,q)$ are arbitrary functions, and $J_{i\nu}(z)$, 
$Y_{i\nu}(z)$ are the Bessel functions of the first and second kind of index 
$i\nu=i|k/q|$ and argument 
\be
z=\sqrt{\lambda}\lambda_s\exp(-q\Phi)/|q|\,,\label{z}
\ee
respectively \cite{AS}.   

Since the space of the solutions (\ref{sol-gen}) is two-dimensional we have 
two sets of (real) orthonormal functions with respect to the inner product 
(\ref{prod-int}) \cite{CD,CDFbh}
\ba
&&\chi^{(1)}_{\nu}(z)=C^{(1)}
\left[e^{-\pi\nu/2}H^{(1)}_{i\nu}(z)+e^{\pi\nu/2}H^{(2)}_{i\nu}(z)\right]
\,,\label{orth-1}\\
&&\chi^{(2)}_{\nu}(z)=iC^{(2)}
\left[e^{-\pi\nu/2}H^{(1)}_{i\nu}(z)-e^{\pi\nu/2}H^{(2)}_{i\nu}(z)\right]
\,,\label{orth-2}
\ea
where 
\be 
C^{(1)}=\sqrt{\nu\cosh(\pi\nu/2)\over 4\sinh(\pi\nu/2)}\,,~~~~~~
C^{(2)}=\sqrt{\nu\sinh(\pi\nu/2)\over 4\cosh(\pi\nu/2)}\,,
\ee
and $H^{(1)}_{i\nu}(z)$ and $H^{(2)}_{i\nu}(z)$ are the Hankel functions of 
the first and second kind respectively \cite{AS}.

Now let us identify the stationary wave functions that correspond to
expanding \pre\ and \post\ phases. We discuss in detail the case $0<q\le 1$ 
leaving at the end of this subsection the discussion of negative values of $q$.

\begin{figure}
\centerline{\epsfxsize=8.0cm
\epsffile{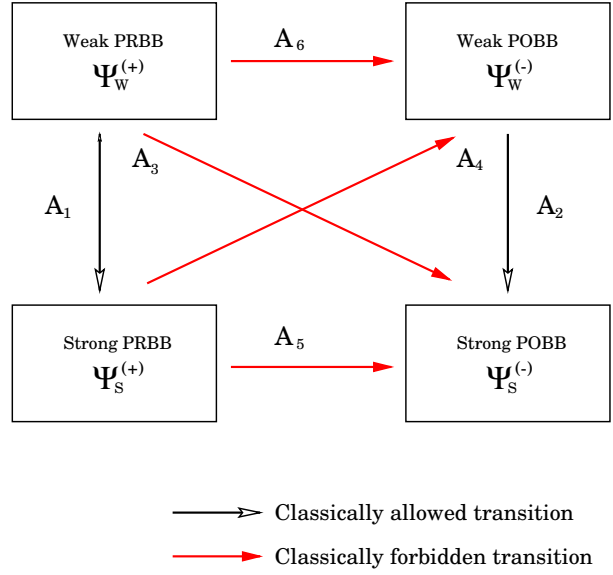}
}
\caption{Different transitions in the weak and strong coupling regimes.}
\label{figuno}
\end{figure}

From (\ref{can-tr-qsc}) and (\ref{H-qsc}) it follows that ${1\over 
2\lambda_s}\Pi_\Theta^2=H_{\rm eff}$. Thus phases that are expanding 
(contracting) are described by eigenstates of the effective Hamiltonian with 
$k>0$ ($k<0$). Expanding wave functions that correspond to \pre\ and \post\ can 
be identified by investigating the asymptotic behaviours of (\ref{orth-1}) and 
(\ref{orth-2}) in the weak and strong coupling regimes. For $z\to\infty$, i.e.\ 
in the weak coupling regime, the wave functions (\ref{orth-1}), (\ref{orth-2}) 
behave as
\ba
&&\chi^{(1)}_{\nu}(z)\approx \sqrt{2\over\pi 
z}\,C^{(1)}\left[e^{i(z-\pi/4)}+e^{-i(z-\pi/4)}\right]\,,\label{chi1-asint}\\
&&\chi^{(2)}_{\nu}(z)\approx i\sqrt{2\over\pi 
z}\,C^{(2)}\left[e^{i(z-\pi/4)}-e^{-i(z-\pi/4)}\right]\,.\label{chi2-asint}
\ea
By applying the momentum operator $\h\Pi_\Phi$ to the linear combinations 
$\chi_{\nu}^{(\pm)}=C^{(2)}\chi^{(1)}_{\nu}\mp iC^{(1)}\chi^{(2)}_{\nu}$ we 
find 
\be
\hat{\Pi}_{\Phi}\chi_{\nu}^{(\pm)}\sim \mp
\lambda_s\sqrt{\lambda}\,e^{-q\Phi}\chi_{\nu}^{(\pm)}\,.
\ee
Thus the wave functions corresponding to \pre\ and \post\ in the weak coupling 
regime are proportional to the linear combinations $\chi_{\nu}^{(\pm)}$ 
respectively. The normalized \pre\ and \post\ wave functions in the weak 
coupling regime are 
\be
\psi^{\rm (\pm)}_W={1\over\sqrt{2\coth(\pi\nu)}}\left[\sqrt{\tanh(\pi\nu/2)}
\chi^{(1)}_{\nu}\mp i\sqrt{\coth(\pi\nu/2)}\chi^{(2)}_{\nu}\right]\,.
\label{wave-weak-pre}\label{wave-weak-prpo}
\ee
By a similar argument we find that the normalized wave functions that 
correspond 
to expanding \pre\ and \post\ phases in the strong coupling regime are 
\be
\psi^{\rm (\pm)}_S={1\over\sqrt{2}}\left[\chi^{(1)}_{\nu}\mp
i\chi^{(2)}_{\nu}\right]\,.\label{wave-strong-prpo}
\ee
Using the two sets of wave functions (\ref{wave-weak-prpo}) and 
(\ref{wave-strong-prpo}) it is possible to compute the amplitudes that 
correspond to the different transitions. They are schematically represented in 
Fig.~(\ref{figuno}), where the amplitudes $A_1\dots A_6$ are given by the 
following expressions
%
%
%
\be
\begin{array}{lll}
&A_1\equiv\left(\psi^{(+)}_S,\psi^{(+)}_W\right)&=~{\displaystyle 
1\over\displaystyle\sqrt{1+e^{-2\pi k/q}}}\,,\\\\
&A_2\equiv\left(\psi^{(-)}_S,\psi^{(-)}_W\right)&=~{\displaystyle 
1\over\displaystyle\sqrt{1+e^{-2\pi k/q}}}\,,\\\\
&A_3\equiv\left(\psi^{(-)}_S,\psi^{(+)}_W\right)&=~-{\displaystyle e^{-\pi 
k/q}\over\displaystyle\sqrt{1+e^{-2\pi k/q}}}\,,\\\\
&A_4\equiv\left(\psi^{(-)}_W,\psi^{(+)}_S\right)&=~-{\displaystyle e^{-\pi 
k/q}\over\displaystyle\sqrt{1+e^{-2\pi k/q}}}\,,\\\\
&A_5\equiv\left(\psi^{(-)}_S,\psi^{(+)}_S\right)&=~0\,,\\\\
&A_6\equiv\left(\psi^{(-)}_W,\psi^{(+)}_W\right)&=~-{\displaystyle 2e^{-\pi 
k/q}\over\displaystyle 1+e^{-2\pi k/q}}\,.
\end{array}
\label{ampl}
\ee
Let us discuss in depth the transition amplitudes (\ref{ampl}). The amplitudes 
$A_1$ and $A_2$ correspond to classically allowed transitions. The relative 
transition probabilities ($P_{S,W}^{(+,+)}\equiv|A_1|^2$, 
$P_{S,W}^{(-,-)}\equiv|A_2|^2$) are 
\be
P_{S,W}^{(+,+)}=P_{S,W}^{(-,-)}={1\over 1+e^{-2\pi k/q}}\,.
\label{prob-prew-pres}
\ee
For $k\to\infty$, i.e.\ in the semiclassical limit, (\ref{prob-prew-pres}) 
becomes 
\be
P_{S,W}^{(+,+)}=P_{S,W}^{(-,-)}\sim 1+O(e^{-2\pi k/q})\,,
\label{semicl-prew-pres}
\ee
in agreement with the classical theory. The amplitudes $A_3$ and $A_4$ describe 
classically forbidden transitions. The relative transition probabilities 
($P_{W,S}^{(+,-)}\equiv|A_3|^2$, $P_{S,W}^{(+,-)}\equiv|A_4|^2$) are
\be
P_{W,S}^{(+,-)}=P_{S,W}^{(+,-)}={e^{-2\pi k/q}\over 1+e^{-2\pi k/q}}\,.
\label{prob-prew-poss}
\ee
In the semiclassical limit (\ref{prob-prew-poss}) becomes
\be
P_{W,S}^{(+,-)}=P_{S,W}^{(+,-)}\sim e^{-2\pi k/q}+O(e^{-4\pi k/q})\,.
\label{semicl-prew-poss} 
\ee
The transition probabilities (\ref{prob-prew-poss}) become highly suppressed 
for $k\gg 1$ where the evolution follows essentially the classical trajectory. 
In the limit $k\to 0$ (\ref{prob-prew-pres}) and (\ref{prob-prew-poss}) become
\be
P_{S,W}^{(+,+)}=P_{S,W}^{(-,-)}\sim 
P_{W,S}^{(+,-)}=P_{S,W}^{(+,-)}\sim {1\over 2}+O(k)\,.
\ee
In the small-$k$ limit quantum effects are significant: the \pre\ (\post) 
phase in the weak coupling regime has the same probability of evolving 
in the \pre\ or in the \post\ phase in the strong coupling regime (and 
viceversa). 

The probability of transition \pre\ $\to$ \post\ in the strong coupling regime 
($P_{S,S}^{(+,-)}=|A_5|^2$) is identically zero. This can be understood looking 
at the asymptotic form of the potential for $\Phi\to\infty$ ($z\to 0$). Indeed,  
for large values of $\Phi$ the potential term in (\ref{schroedinger-stat-qsc}) 
goes asymptotically to zero. As a consequence, \pre\ and \post\ wave functions 
in the strong coupling regime behave asymptotically as free plane waves with 
opposite $\Pi_\Phi$ momentum. Since reflection of free plane waves is forbidden 
the quantum transition from \pre\ to \post\ in the strong coupling regime does 
not take place.

The last and most interesting result is the probability of transition from the 
\pre\ phase in the weak coupling regime to the \post\ phase in the weak coupling 
regime
\be
P_{W,W}^{(+,-)}\equiv\left|A_6\right|^2=4{e^{-2\pi k/q}\over 
\left(1+e^{-2\pi k/q}\right)^2}\,.\label{prob-prew-postw}
\ee
The semiclassical limit of (\ref{prob-prew-postw}) is
\be
P_{W,W}^{(+,-)}\sim 4e^{-2\pi k/q}+O(e^{-4\pi k/q})\,.
\label{semicl-prew-postw}
\ee
For $q=1$ the semiclassical result coincides, apart from a normalisation 
factor, with the `reflection-coefficient' of \cite{GV,GVM}. However, the result 
of  \cite{GV,GVM} should be considered as a ratio between two different 
transition probabilities rather than a transition probability by itself. 
Precisely, the reflection-coefficient defined in \cite{GV,GVM} is 
\be
R\equiv{P_{S,W}^{(-,+)}\over P_{S,W}^{(+,+)}}=e^{-2\pi 
k/q}\,.\label{prob-ratio}
\ee
Note that the (classically forbidden) transition from the strong coupling \pre\ 
phase to the weak coupling \post\ phase is suppressed by a factor $e^{-2\pi 
k/q}$ with respect to the (classically allowed) transition from the strong 
coupling \pre\ phase to the weak coupling \pre\ phase.

Equations (\ref{semicl-prew-pres}), (\ref{semicl-prew-poss}) and  
(\ref{semicl-prew-postw}) give also the asymptotic behaviours for small values 
of $q$ at given $k$. In this case quantum effects are negligible. When $0<q\ll 
1$, the potential in the \Schr\ equation is nearly constant and the \pre\ and 
\post\ solutions are approximated by plane waves of opposite momentum along 
$\Phi$. In this case reflection of waves is highly suppressed.  

A similar analysis can be performed for negative values of $q$. For $q<0$ the 
wave functions that correspond to expanding \pre\ and \post\ phases are
\ba
&&\psi^{(\pm)}_W={1\over\sqrt{2}}\left[\chi^{(1)}_{\nu}\pm i\chi^{(2)}_{\nu}
\right]\,,\nonumber\\
&&\psi^{(\pm)}_S={1\over\sqrt{2\coth(\pi\nu)}}\left[\sqrt{\tanh(\pi\nu/2)}
\chi^{(1)}_{\nu}\pm 
i\sqrt{\coth(\pi\nu/2)}\chi^{(2)}_{\nu}\right]\,,\nonumber\,.
\ea
The amplitudes for the various transitions can be read from~(\ref{ampl}) with 
the substitutions $A_1\leftrightarrow A_2$, $A_5\leftrightarrow A_6$, and $q\ra 
-q$. Now the transition from the weak coupling \pre\ phase to the strong 
coupling \post\ phase is forbidden for negative values of $q$.   

The results of this section show that the probabilities of classically 
forbidden transitions can be expressed, in the semiclassical limit, as power 
series of $e^{-2\pi k/|q|}$. Following \cite{GV,GVM}, from (\ref{Hub}) and 
(\ref{sdil}) we find
\be
\exp(-2\pi k/|q|)=
\exp\left(-\sqrt{4d}\,\pi\Omega_s\over |q|g^2_s\,\lambda^d_s\right)\,,
\label{ist0}
\ee
where $\Omega_s$ is the proper spatial volume and $g_s=e^{\phi_s/2}$ is the 
value of the string coupling when ${\bf H}=1/\lambda_s$. The `istanton-like' 
behaviour of (\ref{ist0}) shows that the probabilities of classically forbidden 
transitions are peaked in the strong coupling regime -- as it has been already 
pointed out in \cite{GV,GVM} -- where all powers of $e^{-2\pi k/|q|}$ have to be 
taken into account. The occurence of this istanton-like behaviour will be 
clarified in the next subsection. 
\subsection{Path integral quantisation}
The string cosmology model that we are considering can also be quantised using
the functional approach. The aim of this subsection is to show how to compute, 
using the path integral formalism, the probability $P_{W,W}^{(+,-)}$ in the 
semiclassical limit. While in the case under investigation the semiclassical 
path integral calculation seems devoid of interest -- we know already the exact 
transition probability (\ref{prob-prew-postw}) -- nevertheless the 
semiclassical calculation is of primary importance if the system cannot be 
quantised exactly. We shall show that the functional approach -- when performed 
appropriately -- reproduces the exact result in the limit of large $k$.  So it 
seems not unreasonable to assume that the semiclassical path integral 
calculation gives a sound approximation of the exact result also for those 
models that are not exactly solvable.  In future, we aim to apply the formalism 
of this subsection to more realistic and interesting models of string cosmology.

The starting point of the functional approach is the path integral in the 
reduced space \cite{HT,IZ}
\be
I=\int_{\Phi(t_1)}^{\Phi(t_2)}\,{\cal D}\Phi\,{\cal 
D}\Pi_\Phi\,\exp\left(iS_{\rm 
eff}[\Phi,\Pi_\Phi]\right)\,,\label{path-gf-qsc}
\ee
where the effective action is given by (\ref{action-eff-qsc}) and 
(\ref{H-eff-qsc}). The transition amplitude $A_6$ is defined by 
(\ref{path-gf-qsc}) where the integral is evaluated on all paths that satisfy 
the boundary conditions
\be
\Phi(-\infty)=-\infty\,,~~~~~~\Phi(\infty)=-\infty\,.
\ee
Since the effective Hamiltonian is quadratic in $\Pi_\Phi$ the integral in 
$\Pi_\Phi$ can be evaluated immediately. We obtain 
\be
I=\int_{\Phi(t_1)}^{\Phi(t_2)}\,{\cal 
D}\Phi\,\exp\left(i\int_{t_1}^{t_2}dt\,{\cal L}_{\rm 
eff}[\Phi,\dot\Phi]\right)\,,\label{path-gf-lagr-qsc}
\label{pathin}
\ee
where the effective Lagrangian is
\be
{\cal L}_{\rm eff}={\lambda_s\over 2}\left(\dot\Phi^2+\lambda
e^{-2q\Phi}\right)\,.\label{gf-lagr-qsc}
\ee
It is advisable to use the variable  $z$ defined in (\ref{z}). Equation 
(\ref{gf-lagr-qsc}) becomes 
\be
{\cal L}_{\rm eff}={1\over 2\lambda_s}\left({\lambda_s^2\over q^2}{\dot 
z^2\over z^2}+q^2z^2\right)\,.\label{gf-lagr-qsc-z}
\ee
Let us first consider the case $0<q\le 1$. The path integral~(\ref{pathin}) 
must be evaluated on all trajectories that satisfy the boundary conditions 
\be
z(-\infty)=\infty\,,~~~~~~z(\infty)=\infty\,.\label{boundary-z}
\ee
The effective Lagrangian (\ref{gf-lagr-qsc-z}) is singular in $z=0$.
So there are no classical solutions describing a (smooth) transition between
\pre\ and \post\ phases (see Sect.\ 2). However, it is possible to construct 
quasi-classical trajectories that satisfy the boundary conditions 
(\ref{boundary-z}) and interpolate between \pre\ and \post\ phases. 

Let us consider the analitical continuation of the variable $z$ to the complex 
plane. The effective Lagrangian is analytical in any point of the complex plane 
$({\rm Re}(z),{\rm Im}(z))$ save for $z=0$. Classically, the transition from 
the weak coupling \pre\ phase to the weak coupling \post\ phase would correspond 
to the trajectory starting at $z=+\infty$, going left along the real axis (\pre\ 
phase, $\dot\Phi>0$), reaching the origin, and finally going right along the 
real axis to $z=+\infty$ (\post\ phase, $\dot\Phi<0$). Clearly, since the 
Lagrangian is singular in $z=0$ a classical continuous and differentiable 
solution does not exist.
%
\begin{figure}
\centerline{\epsfxsize=8.0cm
\epsffile{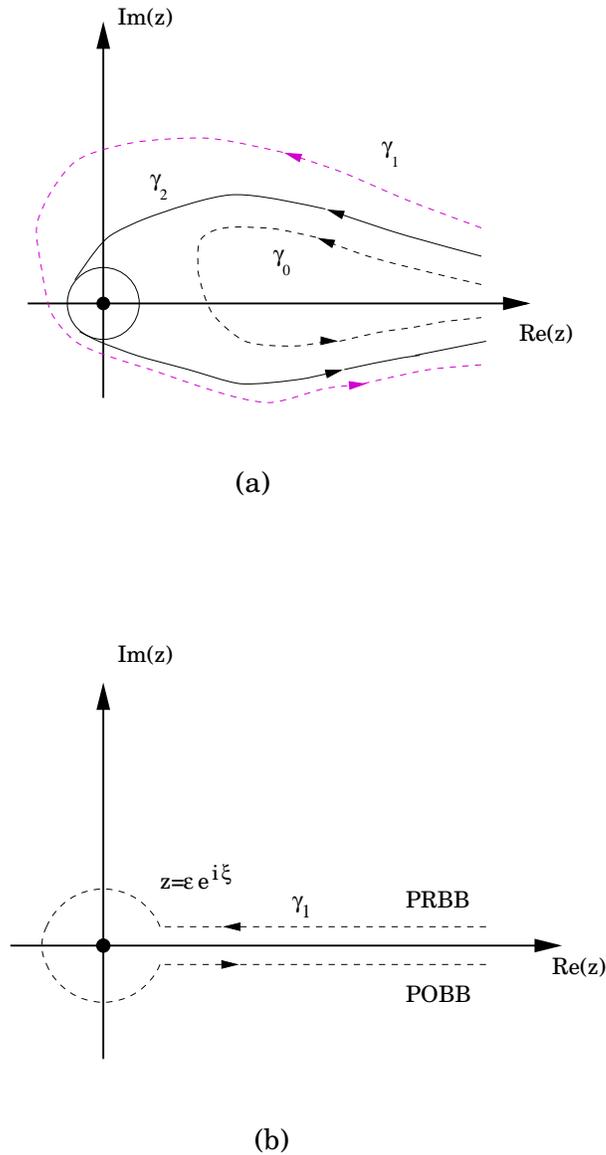}
}
\caption{Contours of integration in the complex $z$-plane.}

\label{figdue}
\end{figure}
%
Now consider generic analytical trajectories in the complex plane that start at 
$Re(z)=\infty$, $Im(z)>0$, and end at $Re(z)=\infty$, $Im(z)<0$ (see 
Fig.\ref{figdue} (a)). We can divide this class of trajectories in three 
(topologically) distinct categories:

{\it i)} Trajectories that do not cross the imaginary axis, i.e.\ trajectories 
that cross the real axis in (at least) one point $z=z_0$, $Re(z_0)>0$, 
$Im(z_0)=0$ (curve $\gamma_0$ in Fig.\ref{figdue} (a));

{\it ii)} Trajectories that cross twice the imaginary axis, i.e.\ trajectories 
that cross once the real axis in $z=z_0$, $Re(z_0)<0$, $Im(z_0)=0$ (curve  
$\gamma_1$ in Fig. \ref{figdue} (a));

{\it iii)} 
Trajectories that cross $2n$ times ($n=2,3...$) the imaginary axis, i.e.\ 
trajectories that cross $n-1$ times the positive real axis and $n$ times the 
negative real axis (curve $\gamma_n$ in Fig. \ref{figdue} (a) for $n=2$); 

Since the action is analytical over the entire complex plane save for $z=0$
trajectories of type $\gamma_0$ can be deformed continuously to a (two-folded) 
trajectory lying entirely on the real positive axis and defined in the interval 
$(Re(z_0),\infty)$. These curves correspond to classical solutions with 
the dilaton field evolving from $\Phi=-\infty$ to a maximum value 
$\Phi=-\ln[qRe(z_0)/\sqrt{\lambda}\lambda_s]/q$ and then decreasing to 
$\Phi=-\infty$. A straightforward calculation shows that the action evaluated 
on this path is identically zero. Since (\ref{gf-lagr-qsc}) is positive 
definite $S_{\rm eff}=0$ can be obtained only by a time reflection, i.e.\ by a 
\pre\ (\post) phase that is covered twice. Therefore, these 
trajectories do not describe transitions from \pre\ to \post\ phases.  

Let us focus attention on trajectories of type $\gamma_1$. They can be deformed
continuously to a trajectory that lies entirely on the real positive axis except
around $z=0$ where the singularity is avoided by the (small) circle
$z=\varepsilon e^{i\xi}$, $\varepsilon\to 0$, $0\le\xi<2\pi$ (see
Fig.\ref{figdue} (b)). This trajectory describes a transition from the weak
coupling \pre\ phase to the weak coupling \post\ phase and corresponds to a
classical solution except in a small region in the strong coupling limit, where
the singularity of the classical solution is avoided by the analytical
continuation in the complex plane. We shall see that the path integral
evaluated on this trajectory gives the leading contribution to the semiclassical
approximation of the transition amplitude $A_6$. Trajectories of type
$\gamma_n$ (with $n\,>\,1$) give contributes of higher order.

It is worth spending a few words on the meaning of the analytical continuation 
of the variable $z$ in the complex plane. Setting $z=Re^{i\xi}$ and using 
(\ref{par}), (\ref{z}) the metric is cast in the form
\be
ds^2=-\left({qR\over\sqrt{\lambda}\lambda_s}\right)^{2/q}e^{2i\xi/q}\mu^2 
dt^2+a^2(t)\,dx_i dx^i\,.\label{line-element}\\
\ee
The signature of (\ref{line-element}) is a function of $\xi$. In particular,
the metric (\ref{line-element}) is real hyperbolic for $\xi=\pi qn$ and real 
Riemannian for $\xi=\pi q(2n+1)/2$, where $n$ is an integer number. Therefore, 
the analitic continuation of Fig.\ref{figdue} (b) can be interpreted as a sort 
of Euclidean analitical continuation in the space of metrics. Any trajectory 
that circles $z=0$ can be considered as an `$n$-instanton' solution (with no 
well-defined signature) labeled by a winding number $n$ that corresponds to the 
number of times that the trajectory wraps around the singularity in $z=0$. In 
the semiclassical limit, the transition amplitude $P_{W,W}^{(+,-)}$ is given by 
the path integral (\ref{path-gf-lagr-qsc}) evaluated on the class of 
$n$-instanton solutions.

Let us consider the contribution to (\ref{path-gf-lagr-qsc}) of the 
one-instanton solution
\be
I_{\rm sc}^{(1)}=C_1\exp\left(iS_{\rm eff}[z_{\gamma_1},\dot 
z_{\gamma_1}]\right)\,,\label{path-gf-semicl}
\ee
where $C_1$ is a normalisation factor and the subscript $\gamma_1$ means 
that the effective action is evaluated along the curve $\gamma_1$ of
Fig.\ref{figdue} (b). For a trajectory with energy $k^2/2\lambda_s$ the 
effective action can be 
cast in the form
\be
S_{\rm eff}=\displaystyle\int_{\gamma_1}\,{dz\over 
z}{1\over\sqrt{z^2+k^2/q^2}}(z^2+k^2/2q^2)\,.\label{path-gf-semicl-k}
\ee
As we do expect, the effective Lagrangian has one isolated singularity in $z=0$ 
(pole of order one). Moreover, for $z\to\infty$ the action 
(\ref{path-gf-semicl-k}) shows a linear divergence. The latter is due to the 
asymptotic behaviour of the \pre\ and \post\ wave functions in the weak 
coupling regime. Indeed, using (\ref{path-gf-semicl}) and 
(\ref{path-gf-semicl-k}) the wave functions corresponding to the \pre\ and 
\post\ phases in the semiclassical approximation are 
\ba
&\psi^{\rm (+)}&\sim\exp{\left[i\int_\infty^z\,{dz'\over 
z'}{1\over\sqrt{z'^2+k^2/q^2}}(z'^2+k^2/2q^2)\right]}\,,\label{semicl1}\\
&\psi^{\rm (-)}&\sim\exp{\left[i\int_z^\infty\,{dz'\over 
z'}{1\over\sqrt{z'^2+k^2/q^2}}(z'^2+k^2/2q^2)\right]}\,.\label{semicl2}
\ea
In the weak coupling  regime ($Im(z)=0$, $z\to\infty$) (\ref{semicl1}) and  
(\ref{semicl2}) behave asymptotically as 
\be
\psi^{\rm (+)}_{z\to\infty}\sim e^{iz}\,,~~~~
\psi^{\rm (-)}_{z\to\infty}\sim e^{-iz}\,,\label{asympt-waves}
\ee
in agreement with the asymptotic behaviour of (\ref{wave-strong-prpo}). 

The integral (\ref{path-gf-semicl-k}) can be made convergent by subtracting the 
asymptotic phase contribution for $z\to\infty$. Then, using the residue 
theorem, we obtain $S_{\rm eff}=\pi ik/q$. The amplitude 
(\ref{path-gf-semicl}) is given by 
\be
I_{\rm sc}^{(1)}=C_1 e^{-\pi k/q}\,.\label{prob-path}
\ee
The semiclassical one-instanton amplitude (\ref{prob-path}) approximates  the 
(exact) result for large values of $k$. This proves the consistency of the 
reduced phase space and path integral quantisation methods. The contribution of 
the $n$-instanton ($n\,>\,1$) to the transition amplitude $A_6$ is
\be
I_{\rm sc}^{(n)}=C_n e^{-\pi nk/q}\,.\label{prob-semicl-n}
\ee
Hence, $n$-instanton terms give higher order contributions in the large-$k$ 
expansion. Equations (\ref{prob-path}) and (\ref{prob-semicl-n}) show that the 
instanton-like dependence (\ref{ist0}) on the string coupling constant of the 
amplitudes that correspond to classically forbidden transitions can be traced 
back to the existence, in the semiclassical regime, of trajectories that 
connect smoothly the \pre\ and \post\ phases. 

Let us conclude this section with two remarks. In the computation of 
(\ref{path-gf-semicl-k}) we have chosen only anticlockwise trajectories (see 
Fig.\ 1 (a-b)). If we considered clockwise paths the residue theorem would give
$S_{\rm eff}=-\pi ik/q$ and the generic contribution to the transition 
amplitude would be 
\be
\tilde I_{\rm sc}^{(n)}=C_n e^{\pi nk/q}\,.\label{prob-semicl-an}
\ee
This result violates -- in the semiclassical limit -- the unitarity
bound. However, there is a simple argument that allows to remove this patology. 
Let us consider the asymptotic behaviours of \pre\ and \post\ wave 
functions in the weak coupling regime. For complex values of $z$ 
(\ref{asympt-waves}) read  
\be
\psi^{\rm (+)}(z\to\infty)\sim e^{iRe(z)-Im(z)}\,,~~~~
\psi^{\rm (-)}(z\to\infty)\sim e^{-iRe(z)+Im(z)}\,.
\label{asympt-waves-complex}
\ee
Since the system must be classical in the weak coupling regime the contribution 
to the path integral of the trajectories that approach the real axis for 
$z\to\infty$ must dominate the contribution of the trajectories with non-zero 
value of $Im(z)$. The above requirement is verified if we integrate along 
anticlockwise trajectories. (In this case the \pre\ and \post\ branches 
are identified by $Im(z)>0$ and $Im(z)<0$ respectively.)

For $q\,<\,0$ the transition amplitude $A_6$ is identically zero. Indeed, 
setting $w=1/z$ the effective Lagrangian~(\ref{gf-lagr-qsc-z}) becomes
\be
{\cal L}_{\rm eff}={1\over 2\lambda_s}\left({\lambda_s^2\over q^2}{\dot 
w^2\over w^2}+{q^2\over w^2}\right)\label{gf-lagr-qsc-w}
\ee
and the path integral~(\ref{path-gf-lagr-qsc}) must be evaluated on
trajectories that satisfy the boundary condition $w(-\infty)=\infty$, 
$w(\infty)=\infty$. The action evaluated on a generic $n$-instanton 
solution is identically zero. Therefore, the semiclassical trajectories do not 
correspond to a transition between \pre\ and \post\ phases. 

\section{Conclusions}
The graceful exit, i.e.\ the transition from the inflationary `pre-big bang 
phase' to the deflationary `post-big bang' phase is a fundamental subject of 
research in (quantum) string cosmology. 

In this paper we have addressed this topic by investigating a special class of 
minisuperspace models that are invariant under scale factor duality 
transformations. Though this particular class of models had been previously 
considered in the literature \cite{GV,GVM,Mah} a deeper discussion was needed. 
Indeed, our analysis clarifies some issues of previous investigations such as 
the meaning of the reflection coefficient and the instanton-like behaviour of 
the \pre\ $\to$ \post\ transition, and provides new interesting results, for 
instance the analysis of the full set of transition amplitudes and the r{\^o}le 
of the semiclassical approximation.

We have shown -- by a concrete example -- that the reduced phase space and the
path integral approaches are extremely powerful techniques of quantisation for a
large class of string cosmology models. The two methods can be applied
straightforwardly to any isotropic, spatially flat, model as long as the latter
is characterized by scale factor duality invariance.  In particular, the
functional method may result very useful when the system cannot be integrated
explicitly, i.e.\ when the \Schr\ equation (or, alternatively, the equivalent
Wheeler-de Witt equation) cannot be solved exactly. Indeed, the calculation of
the (semiclassical) transition amplitude between the \pre\ and \post\ phases in
the weak coupling regime is reduced to a simple evaluation of a definite
integral by means of the residue theorem.  No explicit solutions of the
classical equations of motion nor exact wave functions are needed.

The path integral method makes also clear a couple of other interesting 
features of quantum string cosmology models. First, we have proved that the 
instanton-like nature of the \pre\ $\to$ \post\ transition amplitude 
\cite{GV,GVM} is just a consequence of the presence of the classical 
singularity in the strong coupling regime. Indeed, the mere existence of the 
singularity implies that any semiclassical trajectory gives a $n$-instanton 
contribution to the \pre\ $\to$ \post\ transition amplitude. Second, we have 
clarified the r{\^o}le of the functional form of the dilaton potential in the 
transition process. We have mentioned that the dilaton potential may `mimic' -- 
at the quantum level -- high order corrections to the low-energy effective 
string theory action. The path integral approach shows that the calculation of 
the semiclassical transition amplitude \pre\ $\ra$ \post\ does not require the 
knowledge of the exact functional form of the dilaton potential. The 
semiclassical contribution to the transition amplitude is determined uniquely 
by the behaviour of the dilaton potential in the strong coupling region. Thus 
for any dilaton potential whose asymptotic behaviour for $\Phi\to\infty$ is 
$V\sim e^{a\Phi}$, where $a$ is a real positive parameter, the transition 
amplitude (in the semiclassical approximation) is known.

Let us conclude with an interesting speculation. The transition from the \pre\
phase to the \post\ phase can be (phenomenologically) described by an analytical
continuation of the dilaton field to complex values. We have seen in Sect.\ 3.2
that this analytical continuation can be interpreted in terms of a set of
(complex) metrics with no well-defined signature. This way of looking at an
analytical continued solution as a quantum bridge connecting two classical
hyperbolic spaces has strong resemblance with the semiclassical Euclidean
wormhole picture.  Euclidean wormholes are classical instanton solutions of
gravity-matter systems that (asymptotically) connect two manifolds \cite{CDD}.
They are usually interpreted as tunnelling between the two asymptotic
configurations. In our case the transition from the \pre\ phase to the \post\
phase -- at the semiclassical level -- can be seen precisely as a wormhole-like
effect. Our investigation provides the first example of the calculation of a
wormhole-like tunnelling probability beyond the semiclassical level. This
interpretation is very intriguing and supportes the interesting suggestion that
singularities in the classical domain of physical, hyperbolic solutions in
gravity theories can be avoided by complex solutions joining two spaces, as it
happens in the case that we have discussed here.
\ack
We are very grateful to Maurizio Gasperini, Gabriele Veneziano, and Alex 
Vilenkin for interesting discussions and useful suggestions.

\end{document}